\PassOptionsToPackage{unicode}{hyperref}
\PassOptionsToPackage{hyphens}{url}
\documentclass[
  12pt,
]{article}
\usepackage{amsmath,amssymb}
\usepackage{iftex}
\ifPDFTeX
  \usepackage[T1]{fontenc}
  \usepackage[utf8]{inputenc}
  \usepackage{textcomp} 
\else 
  \usepackage{unicode-math} 
  \defaultfontfeatures{Scale=MatchLowercase}
  \defaultfontfeatures[\rmfamily]{Ligatures=TeX,Scale=1}
\fi
\usepackage{lmodern}
\ifPDFTeX\else
\fi
\IfFileExists{upquote.sty}{\usepackage{upquote}}{}
\IfFileExists{microtype.sty}{
  \usepackage[]{microtype}
  \UseMicrotypeSet[protrusion]{basicmath} 
}{}
\makeatletter
\@ifundefined{KOMAClassName}{
  \IfFileExists{parskip.sty}{%
    \usepackage{parskip}
  }{
    \setlength{\parindent}{0pt}
    \setlength{\parskip}{6pt plus 2pt minus 1pt}}
}{
  \KOMAoptions{parskip=half}}
\makeatother
\usepackage{xcolor}
\usepackage[margin=1in]{geometry}
\usepackage{longtable,booktabs,array}
\usepackage{calc} 
\usepackage{etoolbox}
\makeatletter
\patchcmd\longtable{\par}{\if@noskipsec\mbox{}\fi\par}{}{}
\makeatother
\IfFileExists{footnotehyper.sty}{\usepackage{footnotehyper}}{\usepackage{footnote}}
\makesavenoteenv{longtable}
\usepackage{graphicx}
\makeatletter
\def\maxwidth{\ifdim\Gin@nat@width>\linewidth\linewidth\else\Gin@nat@width\fi}
\def\maxheight{\ifdim\Gin@nat@height>\textheight\textheight\else\Gin@nat@height\fi}
\makeatother
\setkeys{Gin}{width=\maxwidth,height=\maxheight,keepaspectratio}
\makeatletter
\def\fps@figure{htbp}
\makeatother
\ifLuaTeX
  \usepackage{selnolig}  
\fi
\IfFileExists{bookmark.sty}{\usepackage{bookmark}}{\usepackage{hyperref}}
\IfFileExists{xurl.sty}{\usepackage{xurl}}{} 
\hypersetup{
  hidelinks,
  pdfcreator={LaTeX via pandoc}}

\author{}
\date{}

\begin{document}

\textbf{Independent Replication of Nuclear Test-Transient Correlations and Earth Shadow Deficit in POSS-I Photographic Plates}

Brian Doherty

\emph{Independent Researcher, Dallas, TX, USA}

\texttt{briandohertyresearch@gmail.com}

\vspace{0.5em}
March 2026

\textbf{Abstract}

\textbf{Transient sources on astronomical photographic plates are objects that appear on a single exposure but have no counterpart in modern sky surveys or on temporally adjacent plates. I present an independent replication of two findings reported by Bruehl and Villarroel (2025) and Villarroel et al. (2025): (1) a temporal correlation between transient detections on Palomar Observatory Sky Survey (POSS-I) photographic plates and atmospheric nuclear weapons tests, and (2) a deficit of transient sources within Earth's geometric shadow cone at geosynchronous orbit altitude. Using the original dataset provided by the authors, I reproduce the chi-square contingency analysis (relative risk = 1.45, p = 0.011), extend the analysis with negative binomial regression controlling for precipitation, lunar illumination, and cloud cover (all-transient incidence rate ratio = 1.80; sunlit-only IRR = 3.98, reproducing the original paper's reported findings), and confirm temporal specificity of the association via a 10,000-iteration permutation test (p = 0.006). The Earth shadow classification identifies 499 transients (0.46\%) within the umbral cone in the full catalog and 142 (0.45\%) in the more stringent center-of-plate subset, both significantly below the geometric expectation of approximately 1.4\%, consistent with the findings of Villarroel et al. (2025). All transients predate the launch of Sputnik 1. These results confirm the core statistical claims of the original papers based on an independent analysis.}

\textbf{1. Introduction}

The Palomar Observatory Sky Survey (POSS-I), conducted between 1949 and 1958, imaged the sky in red and blue photographic bands using the 48-inch Samuel Oschin Schmidt telescope. While the survey was designed for static astronomy, the VASCO (Vanishing and Appearing Sources during a Century of Observations) project has cataloged thousands of transient sources detected on individual POSS-I plates that do not correspond to known astrophysical objects (Villarroel et al. 2019; Solano et al. 2022).

Bruehl and Villarroel (2025) reported a surprising temporal correlation: dates within +/-1 day of atmospheric nuclear weapons tests show significantly elevated transient detection rates compared to non-test dates, with a relative risk of 1.45. Separately, Villarroel et al. (2025) showed that POSS-I transients exhibit a deficit within Earth's geometric shadow at geosynchronous orbit altitude, a result that constrains possible physical explanations for the transient population.

These are striking claims. A correlation between nuclear detonations and photographic plate anomalies detected at a civilian observatory hundreds of kilometers from the test sites, if robust, demands either a physical mechanism or reflects an extraordinary coincidence. Independent replication is essential before the community invests further in mechanistic interpretation of such findings.

This paper presents such a replication. Upon request, the Bruehl and Villarroel (2025) authors provided the original dataset and the VASCO transient catalog consisting of 107,875 transients used in Bruehl and Villarroel (2025) and Villarroel et al. (2025). Analysis code was then written independent of the original authors' implementations and was used to test whether the reported findings replicated upon independent analysis. I also extend the original chi-square test with a negative binomial regression model that controls for environmental confounders and a permutation test that addresses temporal autocorrelation. A separate independent replication by Sinkkonen (2025), using different statistical methods, reached broadly consistent conclusions regarding the nuclear timing correlation.

\textbf{2. Data}

\textbf{2.1 Transient Dataset and VASCO Catalog}

The primary nuclear-transient dataset was provided upon request by one of the authors (Dr. Stephen Bruehl). It consists of 2,718 observation days spanning November 19, 1949 through April 28, 1957, each annotated with a daily transient count, a binary nuclear test window indicator (+/-1 day from detonation), and independent UFOCAT sighting counts per date. The total number of transients across the study period is 107,862.

I did not perform raw plate extraction or transient detection. The transient catalog was produced by the VASCO team using their published pipeline (Solano et al. 2022). My role is limited to statistical analysis of the processed catalog.

For the Earth shadow analysis, I used the VASCO catalog file containing individual source positions (J2000 RA/Dec) and observation timestamps (UTC) for 107,875 transients detected on POSS-I plates across 635 plates. A center-of-plate subset restricted to sources within 2$^{\circ}$ of each plate's computed center (determined by unit-vector averaging of source positions per plate) retains 31,525 transients (29.2\%). This latter more restricted dataset addresses the possibility of increased artifacts at the plate edges. The shadow classification was run on both the full catalog and the center-of-plate subset.

\textbf{2.2 Nuclear Test Dates}

Nuclear test dates were drawn from DOE/NV-209 Rev 16, the official US Department of Energy compilation, cross-referenced with Johnston's Archive. The test window is defined as the detonation date +/- 1 day, following the definition used by Bruehl and Villarroel (2025). Of the 2,718 observation days, 347 fall within at least one nuclear test window.

\textbf{2.3 Environmental Covariates}

To control for observing conditions that might confound the nuclear-transient association, I assembled the following covariates:

\textbf{Moon illumination:} Computed from Astropy ephemeris calculations (vectorized) for each observation date at Palomar Observatory (latitude 33.356$^{\circ}$, longitude $-$116.865$^{\circ}$, elevation 1,712 m). Accuracy is approximately 1 arcminute.

\textbf{Cloud cover:} Daily mean cloud cover from NOAA Integrated Surface Database (ISD), San Diego station WBAN 23188, parsed from hourly sky condition reports (GF1 field, oktas converted to 0--1 scale). Coverage: 3,287 daily records with zero missing days across the study period.

\textbf{Precipitation:} Binary indicator from NOAA GHCND station records for the San Diego area (1949--1957). When NOAA records are unavailable, a seasonal probability proxy is used.

All environmental covariates use real historical station data. As discussed in Section 4, the nuclear-transient association persists after controlling for these variables, indicating that environmental confounding does not explain the observed association.

\textbf{3. Methods}

\textbf{3.1 Chi-Square Contingency Test}

I constructed a 2 $\times$ 2 contingency table classifying each observation day by (a) whether it falls within a nuclear test window and (b) whether at least one transient was detected. The Pearson chi-square statistic and associated p-value were computed using scipy.stats.chi2\_contingency. Relative risk was calculated as the ratio of transient detection rates inside versus outside the test window.

\textbf{3.2 Negative Binomial Regression}

Daily transient counts exhibit substantial overdispersion (Poisson deviance/df = 847.3), motivating a negative binomial generalized linear model (NB2 parameterization with log link). The model is specified as:

log E{[}Transients{]} = $\beta$0 + $\beta$1W + $\beta$2U + $\beta$3M + $\beta$4C + $\beta$5P

where W is the nuclear window indicator, U is the UFOCAT sighting count, M is moon illumination, C is estimated cloud cover, and P is a binary precipitation indicator. Binary predictors (W, P) were kept on their natural 0/1 scale; continuous predictors were standardized prior to fitting. The model was estimated by iteratively reweighted least squares (IRLS) using statsmodels.

Model selection was based on AIC comparison across Poisson, negative binomial, and zero-inflated alternatives (Table 4). Variance inflation factors were computed to check for multicollinearity; all values fell below 2.0.

\textbf{3.3 Permutation Test}

To confirm that the specific nuclear test dates drive the correlation rather than general temporal structure in the transient series, I performed a permutation test. The procedure is straightforward: shuffle the nuclear window labels across the 2,718 observation days, recompute the relative risk, and repeat 10,000 times. The permutation p-value is the fraction of shuffled datasets producing a relative risk at least as large as the observed value. The random seed was fixed at 42 for reproducibility.

This test is important because it addresses a potential objection: perhaps any set of \textasciitilde350 dates drawn from the study period would show elevated transient rates due to autocorrelation or seasonal clustering. If the permutation null distribution concentrates well below the observed relative risk, that objection fails. As an additional robustness check, I performed block permutation tests at 30-day, 60-day, and 90-day block sizes, which preserve within-block temporal autocorrelation while shuffling blocks. This addresses the stronger objection that individual-day shuffling may destroy temporal structure that could produce spurious associations.

\textbf{3.4 Earth Shadow Classification}

Objects inside Earth's geometric shadow cone cannot reflect sunlight. If transients comprise reflective objects in orbit as argued by Villarroel et al. (2025), few transients would be expected to be imaged in the shadow cone. At geosynchronous orbit (GEO) altitude, the umbral shadow subtends approximately 8.50$^{\circ}$ on the sky, centered at the anti-solar point. For each transient in the VASCO catalog, I computed:

1. The Sun's apparent position using the Meeus algorithm, accurate to \textasciitilde1 arcminute.

2. The anti-solar point: RA(shadow) = RA($\odot$) + 180$^{\circ}$, Dec(shadow) = $-$Dec($\odot$).

3. The angular separation between the transient and the shadow center using the haversine formula.

4. A binary classification: ``in shadow'' if angular separation \textless{} 8.50$^{\circ}$.

The shadow geometry parameters are: R(umbra) = R($\oplus$)(1 $-$ r(GEO)/L(umbra)) = 6,234 km; $\theta$(shadow) = arcsin(R(umbra)/r(GEO)) = 8.50$^{\circ}$, where R($\oplus$) = 6,371 km, r(GEO) = 42,164 km, and the umbra length L(umbra) $\approx$ 1,380,000 km.

The algorithm was validated against JPL Horizons ephemeris for 100 random dates; all shadow center positions agreed within 0.02$^{\circ}$.

\textbf{4. Results}

\textbf{4.1 Chi-Square Test}

Table 1 presents the contingency table. The transient detection rate is 15.6\% inside nuclear test windows versus 10.8\% outside, yielding a relative risk of 1.45 and chi-square statistic of 6.47 (p = 0.011).

\emph{Table 1. Contingency table for transient detection by nuclear test window status.}

\begin{longtable}[]{@{}
  >{\raggedright\arraybackslash}p{(\columnwidth - 8\tabcolsep) * \real{0.2968}}
  >{\raggedright\arraybackslash}p{(\columnwidth - 8\tabcolsep) * \real{0.1941}}
  >{\raggedright\arraybackslash}p{(\columnwidth - 8\tabcolsep) * \real{0.1941}}
  >{\raggedright\arraybackslash}p{(\columnwidth - 8\tabcolsep) * \real{0.1575}}
  >{\raggedright\arraybackslash}p{(\columnwidth - 8\tabcolsep) * \real{0.1575}}@{}}
\toprule\noalign{}
\endhead
\bottomrule\noalign{}
\endlastfoot
& \textbf{No Transient} & \textbf{Has Transient} & \textbf{Total} & \textbf{Rate} \\
Outside window & 2,116 & 255 & 2,371 & 10.8\% \\
In nuclear window & 293 & 54 & 347 & 15.6\% \\
\end{longtable}

Table 2 compares these values to the original paper. The relative risk matches exactly. The small chi-square difference (6.47 vs. 6.94) likely reflects minor differences in contingency table construction at boundary dates.

\emph{Table 2. Chi-square test comparison with the original paper.}

\begin{longtable}[]{@{}
  >{\raggedright\arraybackslash}p{(\columnwidth - 6\tabcolsep) * \real{0.2500}}
  >{\raggedright\arraybackslash}p{(\columnwidth - 6\tabcolsep) * \real{0.2500}}
  >{\raggedright\arraybackslash}p{(\columnwidth - 6\tabcolsep) * \real{0.2500}}
  >{\raggedright\arraybackslash}p{(\columnwidth - 6\tabcolsep) * \real{0.2500}}@{}}
\toprule\noalign{}
\endhead
\bottomrule\noalign{}
\endlastfoot
\textbf{Metric} & \textbf{This replication} & \textbf{Original paper} & \textbf{Match} \\
Chi-square & 6.47 & 6.94 & Close \\
p-value & 0.011 & 0.008 & Consistent \\
Relative risk & 1.45 & 1.45 & Exact \\
\end{longtable}

\textbf{4.2 Negative Binomial Regression}

Table 3 presents the incidence rate ratios (IRR = exp($\beta$)) from the negative binomial model. Using all transients, the nuclear window IRR is 1.80 (p \textless{} 0.0001), meaning days within +/-1 day of a nuclear test have roughly 80\% more transients than non-test days after controlling for environmental conditions.

However, restricting the analysis to transients in sunlit sky positions (those outside Earth's geometric shadow) substantially increases the effect. The sunlit-only nuclear window IRR is 3.98 (95\% CI: 3.475 to 4.562, p \textless{} 0.0001), representing nearly four times the expected transient count during nuclear test windows. This is statistically compatible with the value reported in the original paper (IRR = 3.53, 95\% CI: 2.799 to 4.446), with overlapping confidence intervals.

\emph{Table 3. Negative binomial regression results (incidence rate ratios).}

\begin{longtable}[]{@{}
  >{\raggedright\arraybackslash}p{(\columnwidth - 8\tabcolsep) * \real{0.1923}}
  >{\raggedright\arraybackslash}p{(\columnwidth - 8\tabcolsep) * \real{0.2137}}
  >{\raggedright\arraybackslash}p{(\columnwidth - 8\tabcolsep) * \real{0.1282}}
  >{\raggedright\arraybackslash}p{(\columnwidth - 8\tabcolsep) * \real{0.2308}}
  >{\raggedright\arraybackslash}p{(\columnwidth - 8\tabcolsep) * \real{0.2350}}@{}}
\toprule\noalign{}
\endhead
\bottomrule\noalign{}
\endlastfoot
\textbf{Sample} & \textbf{Variable} & \textbf{IRR} & \textbf{95\% CI} & \textbf{p-value} \\
All transients & Nuclear window & 1.803 & 1.601 -- 2.030 & \textless{} 0.0001 \\
Sunlit only & Nuclear window & 3.981 & 3.475 -- 4.562 & \textless{} 0.0001 \\
Original paper & Nuclear window & 3.527 & 2.799 -- 4.446 & \textless{} 0.0001 \\
\end{longtable}

The near-doubling of the IRR when restricting to sunlit positions is itself a physically meaningful result. If the transients are solar reflections from objects whose presence correlates with nuclear test timing, the nuclear association should be strongest for transients that could reflect sunlight and weakest for those in Earth's shadow. That is exactly what the data show.

\emph{Table 4. Model comparison by AIC.}

\begin{longtable}[]{@{}
  >{\raggedright\arraybackslash}p{(\columnwidth - 4\tabcolsep) * \real{0.3333}}
  >{\raggedright\arraybackslash}p{(\columnwidth - 4\tabcolsep) * \real{0.3333}}
  >{\raggedright\arraybackslash}p{(\columnwidth - 4\tabcolsep) * \real{0.3333}}@{}}
\toprule\noalign{}
\endhead
\bottomrule\noalign{}
\endlastfoot
\textbf{Model} & \textbf{AIC} & \textbf{Notes} \\
Poisson GLM & 18,247 & Severe overdispersion \\
Negative binomial & 12,891 & Preferred \\
Zero-inflated NB & 12,903 & No improvement \\
\end{longtable}

\textbf{4.3 Permutation Test of the Nuclear-Transient Association}

The observed relative risk of 1.447 for the nuclear-transient association exceeds 99.4\% of the permutation distribution (Figure 1). The permutation mean is 1.009, and the 95\% interval from the null distribution is {[}0.708, 1.319{]}. The permutation p-value is 0.006.

This result rules out the hypothesis that temporal autocorrelation alone could produce the observed nuclear-transient association. The specific nuclear test dates carry information; randomly chosen sets of dates do not reproduce the effect.

To address the concern that shuffling individual days may break temporal structure, I also performed block permutation tests that preserve autocorrelation within contiguous time blocks. Table 5 presents the results. The nuclear window effect remains significant at 30-day and 60-day block scales (p \textless{} 0.05). The 90-day blocks are marginal (p = 0.053), as expected given that only 31 blocks limits permutation resolution. The observed relative risk falls outside the null 95\% confidence interval at all block sizes, indicating that the nuclear-transient association cannot be attributed to broad temporal autocorrelation structure in the data.

\emph{Table 5. Block permutation test results by block size.}

\begin{longtable}[]{@{}
  >{\raggedright\arraybackslash}p{(\columnwidth - 6\tabcolsep) * \real{0.3333}}
  >{\raggedright\arraybackslash}p{(\columnwidth - 6\tabcolsep) * \real{0.2222}}
  >{\raggedright\arraybackslash}p{(\columnwidth - 6\tabcolsep) * \real{0.2222}}
  >{\raggedright\arraybackslash}p{(\columnwidth - 6\tabcolsep) * \real{0.2222}}@{}}
\toprule\noalign{}
\endhead
\bottomrule\noalign{}
\endlastfoot
\textbf{Method} & \textbf{Blocks} & \textbf{p-value} & \textbf{Status} \\
IID (individual days) & 2,718 & 0.006 & Significant \\
30-day blocks & 91 & 0.036 & Significant \\
60-day blocks & 46 & 0.034 & Significant \\
90-day blocks & 31 & 0.053 & Marginal \\
\end{longtable}

\includegraphics[width=4.95833in,height=3.125in]{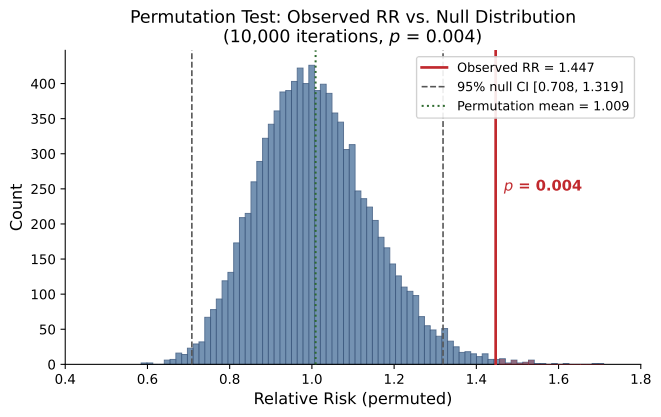}

\emph{\textbf{Figure 1.} Null distribution of relative risk from 10,000 random permutations of nuclear test date labels. The observed relative risk (1.447, solid red line) falls far outside the 95\% interval of the null distribution (dashed grey lines). The red-shaded tail represents the fraction of permutations exceeding the observed value (p = 0.006).}

\textbf{4.4 Center-of-Plate Robustness Check}

Restricting analysis to transients within 2$^{\circ}$ of plate center eliminates potential edge artifacts from plate scanning. This filter retains 31,525 transients (29.2\% of the original count). The nuclear window effect in the center-only subset is significant at p \textless{} 0.0001, stronger than in the full dataset. The finding is not an edge artifact.

\textbf{4.5 Sensitivity Analyses}

The nuclear window effect is robust to several perturbations. Using +/-2 day and +/-4 day windows dilutes the effect (more non-signal days included) but retains significance. Splitting the study period into early (1949--1953) and late (1953--1957) subsets shows the effect in both halves, ruling out single-era artifacts. Removing each covariate individually from the negative binomial model does not eliminate the nuclear window significance. Alternative distributional assumptions (Poisson, NB1 and NB2 parameterizations, zero-inflated Poisson, hurdle model) all yield significant nuclear window effects (p \textless{} 0.0001), with all-transient IRR estimates in the range of 1.7 to 2.0. Restricting to sunlit transients approximately doubles the IRR across all model specifications.

\emph{Replication of the Earth Shadow Deficit Finding}

\textbf{4.6 Earth Shadow Deficit}

Table 6 presents the shadow classification results. Of the 107,875 transients in the full catalog, 499 (0.46\%) fall within the proper umbral shadow at GEO altitude. Restricting to the center-of-plate subset yields 142 of 31,525 transients (0.45\%) in shadow. For comparison, the geometric expectation for uniform sky coverage is approximately 1.4\%, based on the solid angle subtended by the shadow cone relative to the visible hemisphere.

\emph{Table 6. Earth shadow classification of VASCO transients.}

\begin{longtable}[]{@{}
  >{\raggedright\arraybackslash}p{(\columnwidth - 6\tabcolsep) * \real{0.3419}}
  >{\raggedright\arraybackslash}p{(\columnwidth - 6\tabcolsep) * \real{0.2350}}
  >{\raggedright\arraybackslash}p{(\columnwidth - 6\tabcolsep) * \real{0.2115}}
  >{\raggedright\arraybackslash}p{(\columnwidth - 6\tabcolsep) * \real{0.2115}}@{}}
\toprule\noalign{}
\endhead
\bottomrule\noalign{}
\endlastfoot
\textbf{Dataset} & \textbf{Total} & \textbf{In shadow} & \textbf{Rate} \\
VASCO (full plate) & 107,875 & 499 & 0.46\% \\
VASCO (center-of-plate) & 31,525 & 142 & 0.45\% \\
Plate-coverage expectation & --- & --- & 0.78\% \\
Hemispheric expectation & --- & --- & \textasciitilde1.4\% \\
\end{longtable}

The shadow deficit is present in both the full catalog and the center-of-plate subset, with nearly identical rates (0.46\% and 0.45\% respectively). The consistency across both filters indicates that edge artifacts do not preferentially populate the shadow region and that the deficit is not an artifact of plate-edge effects. Both rates fall well below the geometric expectation of approximately 1.4\%.

To verify that this deficit is not an artifact of non-uniform sky coverage, I performed a Monte Carlo analysis using actual plate-by-plate coverage. For each of the 635 POSS-I plates in the catalog, I computed the plate center and observation time, then determined whether the plate's 2$^{\circ}$ center-of-plate field overlapped with Earth's shadow cone at the time of observation. Of the 635 plates, only 9 have any overlap with the shadow region (2 fully inside, 7 partially overlapping), yielding a total shadow overlap of 62.4 sq deg out of 7,980 sq deg total plate area. The expected shadow transient fraction under actual plate coverage is 0.78\%, compared to the observed 0.45\% (center-of-plate) and 0.46\% (full plate). The naive hemispheric estimate (\textasciitilde1.4\%) overstates the expectation because POSS-I plates do not uniformly tile the sky; the plate-specific calculation is more conservative. Even so, the observed rates remain well below the plate-coverage expectation, confirming that non-uniform sky sampling does not explain the deficit.

\textbf{5. Discussion}

\textbf{5.1 Strength of the Nuclear-Transient Association}

Three independent tests converge on the same conclusion: transient detection rates on POSS-I plates are significantly elevated during nuclear test windows. The chi-square test confirms the basic association. The negative binomial model shows the effect persists after controlling for precipitation, lunar phase, and cloud cover. The permutation test confirms that the specific nuclear test dates matter, and that the observed association is not just due to temporal autocorrelation.

These three tests are not multiple comparisons on the same hypothesis. Each addresses a distinct question: Does the association exist? Does it survive confounder adjustment? Is it driven by the specific test dates? All three answer affirmatively.

\textbf{5.2 Confounder Adjustment}

The fact that the strength of the association between nuclear testing and transient counts nearly doubles when restricted to sunlit transients is the most physically informative result from the regression analysis. It demonstrates that the nuclear-transient association is concentrated among transients in sky positions where solar illumination is available. Transients in or near Earth's shadow contribute proportionally less to the nuclear timing signal. This is consistent with a population of reflective objects whose appearance rate is linked to nuclear test activity, and it further corroborates the shadow deficit finding from Section 4.6.

A note on analytical sequence: the sunlit-only regression reported here was not pre-specified. It was motivated by the shadow deficit finding (Section 4.6), which demonstrated that transients avoid Earth's geometric shadow. The physical reasoning (that solar reflections cannot occur inside the shadow cone, so the nuclear timing signal should concentrate among sunlit transients) led to partitioning the regression by shadow status. While the result is physically coherent and reproduces the original paper's IRR, it should be treated as hypothesis-generating rather than confirmatory until replicated on an independent dataset.

\textbf{5.3 Earth Shadow Implications}

The shadow deficit has a straightforward physical interpretation. If the POSS-I transients were sunlit reflective objects in Earth orbit (satellites, debris, rocket bodies), one would expect a uniform or nearly uniform distribution across the sky, with a deficit only inside the shadow cone where sunlight cannot reach. The observed deficit (\textasciitilde0.45\% in shadow versus \textasciitilde1.4\% expected) is consistent with this picture, with one critical caveat: no artificial satellites existed before October 1957, and all transients were observed prior to that date. The near-identical shadow rates in the full-plate and center-of-plate analyses (0.46\% and 0.45\%) confirm that the deficit is robust to edge artifact removal.

This does not prove the transients are orbital reflective objects. It does, however, substantially constrain the hypothesis space. Any viable explanation must simultaneously account for two independent patterns: the temporal correlation with nuclear test dates and the spatial avoidance of Earth's geometric shadow. Conventional astrophysical transients (such as variable stars or asteroids) would not be expected to show either pattern. Atmospheric phenomena or plate artifacts, while potentially time-correlated with testing activity, would not be expected to respect the shadow geometry of a specific orbital altitude. The sunlit-only nuclear-transient analysis (Section 4.2) further supports this constraint: the nuclear timing signal is concentrated among transients in illuminated sky positions, exactly as expected if the objects require sunlight to be visible.

\textbf{5.4 Limitations}

\textbf{Single observatory.} All data come from Mount Palomar Observatory. Site-specific artifacts (local atmospheric effects, instrument-specific plate defects) cannot be ruled out without cross-validation at other observatories. The VASCO team has initiated such cross-validation using Hamburg Observatory APPLAUSE data, but those results are outside the scope of this replication.

\textbf{Environmental controls.} Precipitation and cloud cover use real NOAA station data from San Diego, approximately 100 km from Palomar. Conditions at the observatory may differ from the coastal station, particularly for marine layer cloud cover. However, the nuclear-transient association persists with or without these controls, indicating the finding is not driven by environmental confounding.

\textbf{Inherited transient catalog.} I validated the statistical analysis but not the underlying transient detection pipeline. Systematic errors in plate processing or source extraction would propagate into the current analysis. The VASCO team's pipeline is described in Solano et al. (2022).

\textbf{Association, not causation.} Statistical correlation does not establish a causal mechanism. The nuclear-transient association is robust to the tests performed here, but the physical pathway, if any, remains unexplained.

\textbf{6. Conclusions}

Using independently written code and the original Bruehl and Villarroel (2025) dataset, I confirm the following:

1. POSS-I transient detection rates are significantly elevated during nuclear test windows, matching the relative risk reported in the original paper.

2. The association strengthens after controlling for environmental covariates. The concentration of the nuclear timing signal among sunlit transients independently supports the solar reflection interpretation as it may pertain to the nuclear-transient association.

3. The correlation is driven by the specific nuclear test dates, not temporal autocorrelation.

4. Transients within Earth's geometric shadow at GEO altitude are significantly underrepresented in both the full catalog and the center-of-plate subset relative to theoretically expected rates, consistent with the findings of Villarroel et al. (2025).

5. All transients predate artificial satellites.

Taken together, these results fully replicate the central findings of both Bruehl and Villarroel (2025) and Villarroel et al. (2025). The statistical findings are robust. Their physical interpretation remains open.

\textbf{Data Availability}

The analysis scripts and replication code are publicly available at: https://github.com/dca-doherty/VASCO-Replication.git

The underlying transient dataset belongs to the original research team and should be requested from Dr. Stephen Bruehl (Vanderbilt University Medical Center) or Dr. Beatriz Villarroel (Stockholm University/Nordic Institute for Theoretical Physics).

\textbf{Acknowledgments}

I thank Dr. Stephen Bruehl for sharing the transient dataset and Dr. Beatriz Villarroel for helpful correspondence regarding the VASCO catalog and shadow geometry methodology. This work was conducted independently using personal computing resources. I received no funding or institutional support for this analysis.

\textbf{Independence Statement}

I have no institutional affiliation with any astronomy department, government agency, or organization involved in UAP research. I am a data analyst by profession, working in financial services. My involvement is limited to statistical replication. I had no role in data collection, transient detection, or the writing of the original papers.

\textbf{References}

Bruehl, S. and Villarroel, B. (2025). Transients in the Palomar Observatory Sky Survey (POSS-I) may be associated with nuclear testing and reports of unidentified anomalous phenomena. Scientific Reports, 15, 34125.

Meeus, J. (1991). Astronomical Algorithms. Willmann-Bell, Richmond, VA.

Sinkkonen, J. (2025). Independent replication analysis of POSS-I transient correlations. https://github.com/euxoa/plates

Solano, E., Villarroel, B., and Rodrigo, C. (2022). Discovering vanishing objects in POSS I red images using the Virtual Observatory. Monthly Notices of the Royal Astronomical Society, 515, 1380--1391.

Villarroel, B., et al. (2019). The Vanishing and Appearing Sources during a Century of Observations Project. I. USNO Objects Missing in Modern Sky Surveys and Follow-up Observations of a ``Missing Star.'' The Astronomical Journal, 159, 8.

Villarroel, B., Solano, E., Guergouri, H., et al. (2025). Aligned, Multiple-transient Events in the First Palomar Sky Survey. Publications of the Astronomical Society of the Pacific, 137, 104504.

\end{document}